\documentclass{article}

\usepackage{arxiv}

\usepackage[utf8]{inputenc} 
\usepackage[T1]{fontenc}    
\usepackage{hyperref}       
\usepackage{url}            
\usepackage{booktabs}       
\usepackage{amsfonts}       
\usepackage{nicefrac}       
\usepackage{microtype}      
\usepackage{lipsum}		
\usepackage{graphicx}
\usepackage{natbib}
\usepackage{doi}
\usepackage{bm}
\usepackage[varvw]{newtxmath}       
\usepackage{amsmath}
\mathchardef\mhyphen="2D 
\usepackage{booktabs}
\usepackage{multirow}
\usepackage{comment}

\title{Reliability-Based Robust Design Optimization Method for Engineering Systems with Uncertainty Quantification}

\author{ {Richa ~Verma} \\
	Department of Electrical Engineering\\
	 Indian Institute of Technology, Delhi \\
	Hauz Khas, NewDelhi-110 016\\
	\And
         {Dinesh ~Kumar } \\
	Department of Mechanical Engineering\\
	University of Bristol\\
	Bristol BS8 1TR, UK \\
\And
	{Kazuma ~obayashi} \\
	Department of Nuclear Engineering and Radiation Science\\
	Missouri University of Science and Technology\\
	Rolla, MO 65409, USA \\
 \And
      {Syed ~Alam} \\
	Department of Nuclear Engineering and Radiation Science\\
	Missouri University of Science and Technology\\
	Rolla, MO 65409, USA \\
 }

\renewcommand{\shorttitle}{\textit{Handbook of Smart Energy Systems} }

\hypersetup{
pdftitle={A template for the arxiv style},
pdfsubject={q-bio.NC, q-bio.QM},
pdfauthor={David S.~Hippocampus, Elias D.~Striatum},
pdfkeywords={First keyword, Second keyword, More},
}

\begin{document}
\maketitle

\begin{abstract}
Robust optimization is a method for optimization under uncertainties in engineering systems and designs for applications ranging from aeronautics to nuclear. In a robust design process, parameter variability (or uncertainty) is incorporated into the engineering systems’ optimization process to assure the systems’ quality and reliability. This chapter focuses on a robust optimization approach for developing robust and reliable advanced systems and explains the framework for using uncertainty quantification and optimization techniques. For the uncertainty analysis, a polynomial chaos-based approach is combined with the optimization algorithms MOSA (Multi-Objective Simulated Annealing), and the process is discussed with a simplified test function. For the optimization process, gradient-free genetic algorithms are considered as the optimizer scans the whole design space, and the optimal values are not always dependent on the initial values.
\end{abstract}


\section{Introduction}

These days, mathematical modeling and simulations are used for designing engineering equipment in almost all industries. Numerical optimization algorithms are incorporated into the modeling and simulation process to design optimal performance engineering equipment. The main goal of the design process is to ensure all product specifications consider variability and uncertainty in the design response \cite{taylor1997introduction, ghanem2017handbook}. A class of uncertainties originating from system variability, such as boundary conditions used in the mathematical modeling, input parameters (temperature, load, fluid, and material densities), etc, are called aleatory uncertainties. Variability, uncertainty, and tolerance must be incorporated into the system design process to assure the quality and reliability of the design. Therefore, optimal product parameters where the system behavior is robust despite unavoidable variability must be evaluated. In the early design process of a product, a robust optimization process is a new, cost-effective and innovative approach that allows to include uncertainties in the optimization process. From the modeling perspective, it is also necessary to understand the engine performance near or around the optimal point in the design process. For example, in designing a jet engine with maximum engine efficiency, one needs to include variations in operating conditions, geometry roughness/ tolerances, and uncertainties in the material properties as the engine will not operate strictly at the optimal points but at nearby points also. Hence, the goal is to maximize efficiency while ensuring the engine doesn't fail the emission requirements, and this process ensures that engine efficiency is also acceptable for the known uncertainties.

The classical method for optimization process or deterministic optimization doesn't provide an accurate picture of the optimization process of the system under consideration for the input parameters not known properly or are uncertain. In practice, the optimized solution without considering environmental changes or manufacturing uncertainties may provide very catastrophic results. Several stochastic methods have been developed in the last few decades to tackle uncertainties in system parameters or design variables. In mechanical engineering, the aim of structural optimization is to design a very high-performance, robust, and reliable system. Nondeterministic or robust optimization approaches are usually divided into reliability-based design optimization (RBDO) and robust design optimization (RDO). In RBDO, one seeks an optimal design with a given probability of failure. In RDO, the main aim is to find an optimal design with reduced uncertainties of the system objectives. Hence, the optimization problem can be a multi-objective optimization. The combined technique can be called reliability-based robust design optimization (RRBO) for a robust and reliable design. In RRBO, both RDO and RBDO algorithms are employed to find a robust optimal point by adding some constraints to the probability of system responses \cite{kumar2022multi,kumar2019influence,kumar2020nuclear,kumar2020efficient}.

Several RRDO methods \cite{sandgren2002robust, kumar2019combination, ben2002robust, kumar2020uncertainty} have been reported in recent years. Stochastic design optimization using Monte Carlo (MC) based uncertainty quantification is the most simplified method. It can estimate the probability of failure and the objectives' moments of the system directly, but a large number of simulation points are needed for accurate stochastic responses. It makes the Monte Carlo use in the stochastic design process very time-consuming. Many simulation samples are needed to ensure accuracy, making MC-based robust optimization very time-consuming for implementation when numerical simulations for real-life engineering test cases are involved. Alternatively, a more recent uncertainty quantification methods based on polynomial chaos \cite{najm2009uncertainty, xiu2002wiener} is used by numerous researchers. For optimization, two kinds of algorithms are usually used: gradient-based and gradient-free genetic algorithms. In polynomial chaos-based robust optimization, the polynomial chaos method is used to estimate the mean and standard deviations of the objective functions at each iteration point of the design process.

\section{Optimization Under Uncertainty}

Robust optimization is a process of considering uncertain parameters with their variabilities. Instead of finding the optimal point for constant inputs, extracting the uncertain input parameters and their variability, and then finding the optimal point that is robust to those parameters' variabilities. It becomes challenging to have a stable optimal point in a design process that doesn't result in instability or failure, significantly when the optimal response varies due to system uncertainties.

A robust design is considered less sensitive or insensitive to input parameters' uncertainties. The robust optimization process aims to enhance a system's quality by optimizing its performance where input uncertainties are not ignored in the design process. For robust optimization analysis, there is no universal formulation. The designer can set robust design conditions, parameters, and constraints depending on the customers' requirements. A robust problem usually contains the objective function's mean and standard deviations where the objective function's standard deviation or uncertainty is minimized. Depending on the situation, constraints are also added for reliability analysis.

\begin{figure}[!htbp]
\centering
\includegraphics[width=0.8\textwidth]{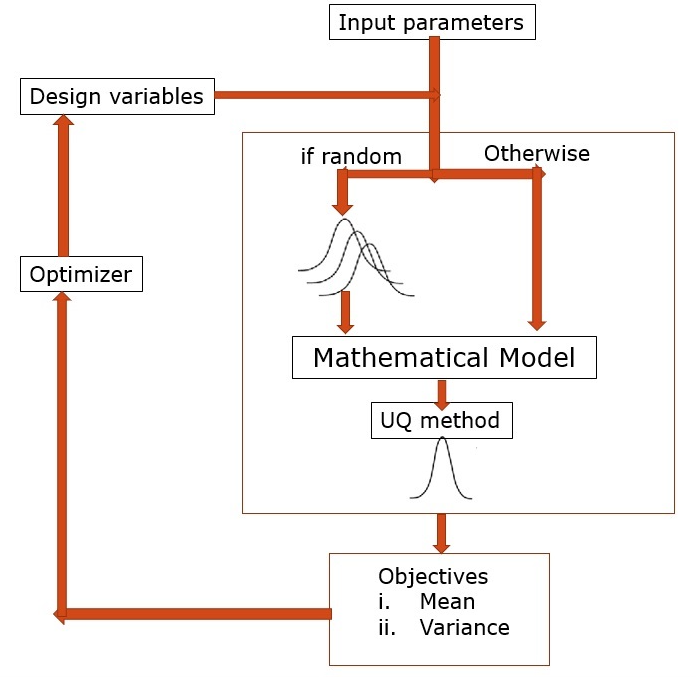}
\caption{Optimization under uncertainties process}
\label{RO}
\end{figure}

Consider a robust optimization problem as:

\begin{equation*}
\label{eq:obj}
   minimize: \ \   V(\bm{x}) = [\mu_V(\bm{x}), \sigma_V(\bm{x})],
\end{equation*}

\begin{equation}
\label{eq:const}
subject \ \ to: \ \   U_i(\bm{x}) > 0
\end{equation}

where $\mu_V(\bm{x}$ and $\sigma_V(\bm{x})$ are the mean and standard deviation of the objective function. The robust optimization problem is defined as a multi-objective problem in the above formulation. A multi-objective problem usually has multiple solutions. These solution points are called points on the Pareto front where based on the customers' demand and criteria, the designer can choose the best design. Figure \ref{RO} depicts a typical flowchart for optimization under uncertainties.

A very long computation time and computational resources are the main challenges in robust optimization. Hence, it's only possible to conduct a robust optimization process on small components of engineering designs due to technical feasibility. Therefore, it is necessary to use efficient optimization algorithms and uncertainty quantification within the process. Alternatively, its also advisable to use surrogate models to speed up the process.

\section{Test Case and Analysis}

\begin{figure}[!htbp]
\centering
\includegraphics[scale=0.55]{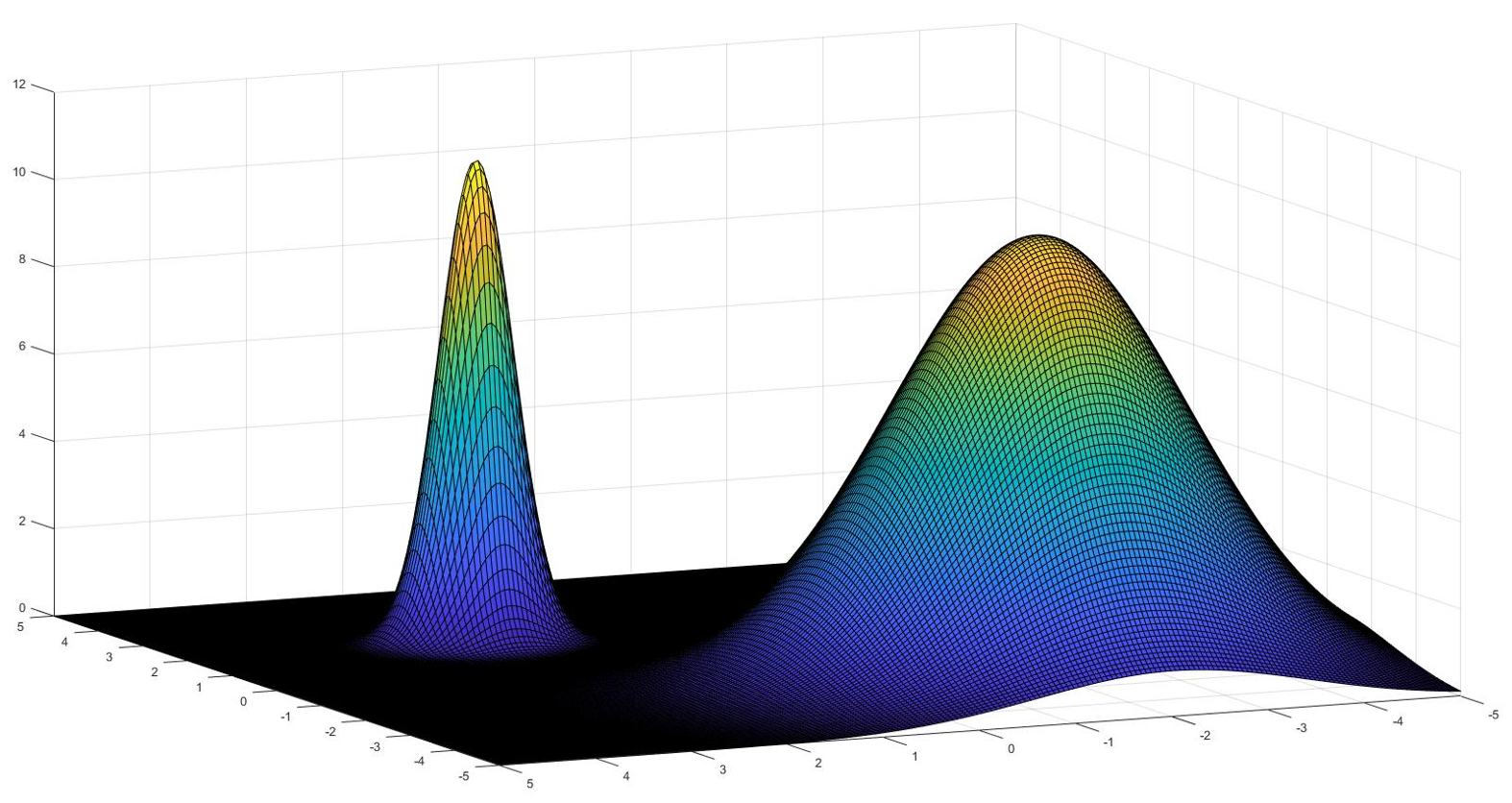}
\caption{Two peak Gaussian function $f(x,y) = k_1 e^{-\frac{(x-x_1)^2+(y-y_1)^2}{2a_1^2}} + k_2 e^ {-\frac{(x-x_2)^2+(y-y_2)^2}{2a_2^2}}$.}
\label{TestFunc}
\end{figure}

In this section, we consider a two peaks function (shown in Fig. \ref{TestFunc}) as described below for the optimization under uncertainties analysis.

\begin{equation}
\label{eq:TestFunction}
   f(x,y) = k_1 e^{-\frac{(x-x_1)^2+(y-y_1)^2}{2\sigma_1^2}} + k_2 e^ {-\frac{(x-x_2)^2+(y-y_2)^2}{2\sigma_2^2}} 
\end{equation}

The above function is a weighted sum of two Gaussian functions, The function have two peaks at two different locations $(x_1=2,y_1=2)$ and $(x_2=-2,y_2=-2)$. For the given coefficients $(k_1=11, \sigma_{1}=0.5)$ and $(k_{2}=10, \sigma_{2}=2)$, the height of the peak at location $(x_1=2,y_1=2)$ is equal to 11 and for $(k_1=10, \sigma_{1}=2)$, the peak at location $(x_2=-2,y_2=-2)$, the height is 10. Hence, it is very clear that the maximum of this function is equal to 11 at the location point $(x_2=2,y_2=2)$. In this section, we will see how the impact of uncertainties finds a maximum point that is robust with respect to uncertainties. 

\begin{figure}[!htbp]
\centering
\includegraphics[scale=0.4]{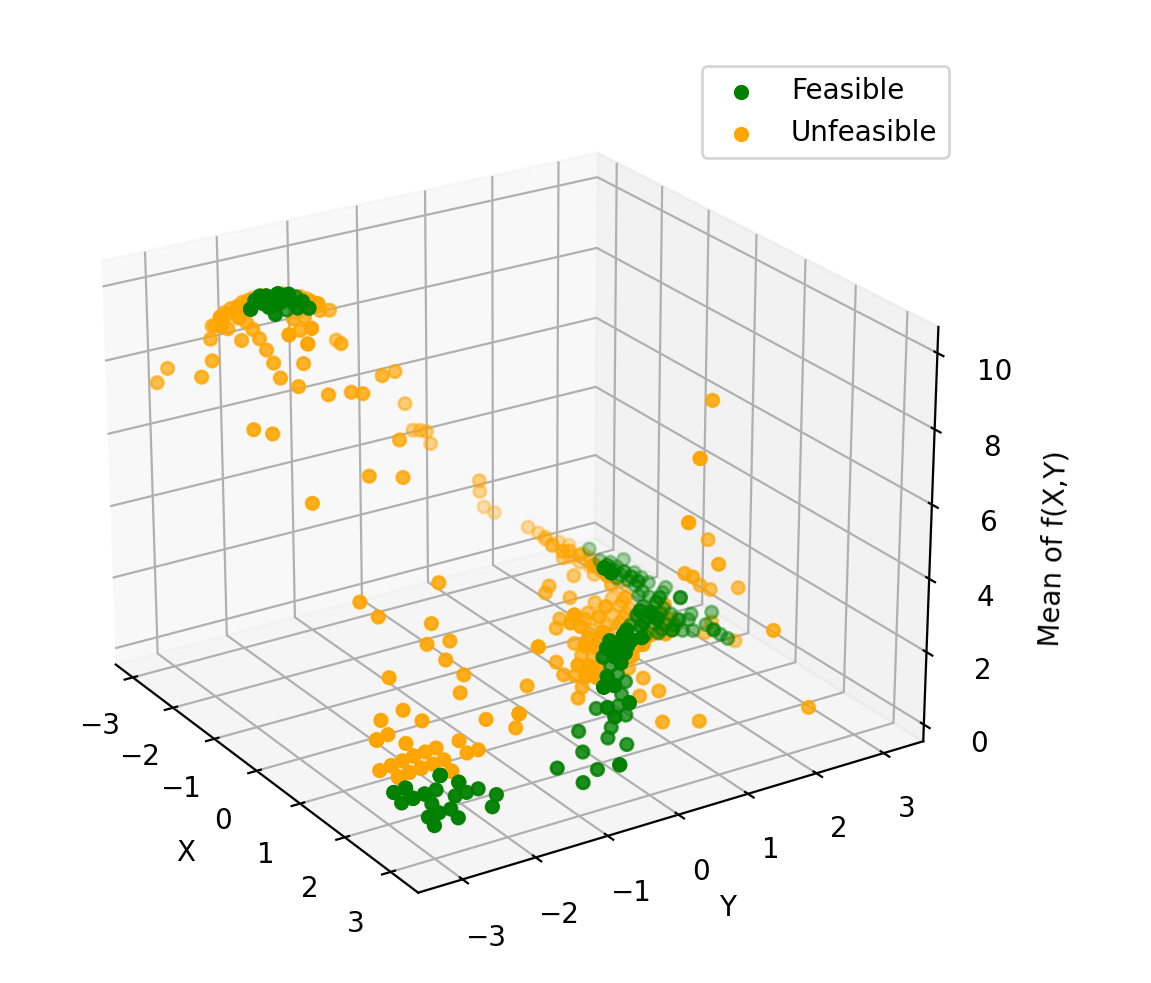}
\caption{Result of multi-objective robust optimization computed with MOSA method \cite{ulungu1999mosa}. As the constraint, a standard deviation of less than 0.1 is imposed. The global maxima appears around $(x=-2, y=-2)$.}
\label{fig:robust}
\end{figure}

\begin{figure}[!htbp]
\centering
\includegraphics[scale=1]{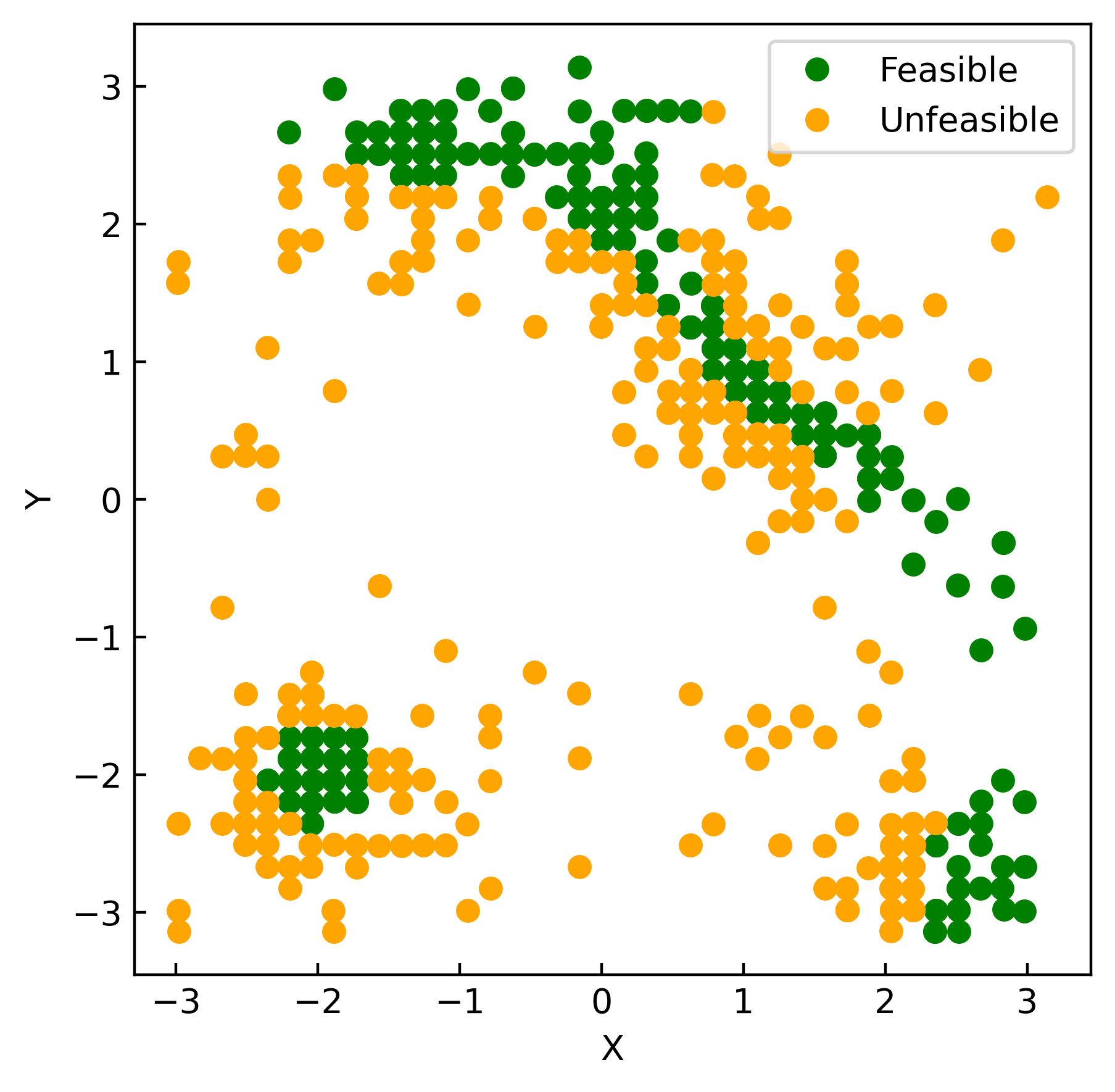}
\caption{Scatter plot for the robust design parameters. The global maxima appears around $(x=-2, y=-2)$. The local maxima $(x=2, y=2)$ is treated as an unfeasible design in the robust optimization.}
\label{fig:robust_params}
\end{figure}

In robust optimization problems, stochastic input variables are required. Based on this concept, $x$ and $y$ must be set. In this study, we impose a normal distribution with a standard deviation of 0.1 for the variables. It means that extra investigating points following the distribution are generated around each robust design. Latin hypercube sampling (LHS) is selected as the sampling method and it generates 50 points around the robust design. In order to find a stable solution, we set the following objectives for the function: (1) maximizing the mean value and (2) obtaining the lowest standard deviation (less than 0.1). The optimization process is implemented using multi-objective simulated annealing (MOSA) \cite{ulungu1999mosa, metropolis1953equation, kirkpatrick1983optimization} and 600 evaluations are performed.

Fig. \ref{fig:robust} shows all the computed designs with the 3D scatter plot. The feasible designs are expressed as green markers and the unfeasible designs as orange. In addition, Fig. \ref{fig:robust_params} represents the 2D scatter plot of the computed design parameters. From Fig. \ref{fig:robust} and Fig. \ref{fig:robust_params} it is revealed that the local maxima $(x=2, y=2)$ does not appears in the robust optimization.

From the nature of the test case, it can be easily predicted that for the optimization case under uncertainties in the design parameters $(x,y)$, the maxima of the mean function will correspond to the peak at the location $(x_2=-2,y_2=-2)$. This peak is not global maxima but under uncertainties when the optimization algorithm looks for a peak that is the most stable with respect to uncertainties. That means, in the nearby points, the function doesn't drop a lot and hence it is robust to the input uncertainties. Hence, the numerical results verify the predicted results.

\section{Conclusions}

In this chapter, first, a brief introduction to the optimization under uncertainty process is described. Further, a test function is considered to explain the process. A simplified test case containing two peaks is considered. One sharp peak, where the maxima of the function take place is the global maxima. The other peak is a bit smaller and is relatively less sharp. The smaller peak is robust with respect to uncertainties. Hence, it corresponds to the maxima for robust optimization. Multi-objective simulated annealing (MOSA) is considered here for the optimization analysis. For uncertainty quantification, the polynomial chaos approach is used. The results from robust optimization are logically discussed with the optimization results with no uncertainties. 

\section*{Acknowledgement}
The computational part of this work was supported in part by the National Science Foundation (NSF) under Grant No. OAC-1919789.

\bibliographystyle{unsrtnat}
\bibliography{references}  






\end{document}